# Monolayer Vanadium-doped Tungsten Disulfide: A Room-Temperature Dilute Magnetic Semiconductor

Fu Zhang[1,2], Boyang Zheng[3], Amritanand Sebastian[4], Hans Olson[5], Mingzu Liu[2,3], Kazunori Fujisawa[2,3,6], Yen Thi Hai Pham[7], Valery Ortiz Jimenez[7], Vijaysankar Kalappattil[7], Leixin Miao[1], Tianyi Zhang[1], Rahul Pendurthi[4], Yu Lei[1,2], Ana Laura Elías[2,3,8], Yuanxi Wang[2,3,9], Nasim Alem[1], Patrick E. Hopkins[5], Saptarshi Das[1,4], Vincent H. Crespi[3,9*], Manh-Huong Phan[7*], Mauricio Terrones[1,2,3,10*]

[1]Department of Materials Science and Engineering, The Pennsylvania State University, University Park, PA 16802, USA
[2]Center for 2- Dimensional and Layered Materials, The Pennsylvania State University, University Park, PA 16802, USA
[3]Department of Physics, The Pennsylvania State University, University Park, PA 16802 USA
[4]Engineering Science and Mechanics, The Pennsylvania State University, University Park, PA 16802, USA
[5]Department of Mechanical and Aerospace Engineering, University of Virginia, Charlottesville, VA 22904, USA
[6]Research Initiative for Supra-Materials, Shinshu University, 4-17-1 Wakasato, Nagano 380-8553, Japan
[7]Department of Physics, University of South Florida, Tampa, Florida 33620, USA
[8]Department of Physics, Binghamton University, Binghamton, NY 13902, USA
[9]2D Crystal Consortium, The Pennsylvania State University, University Park, PA 16802, USA
[10]Department of Chemistry, The Pennsylvania State University, University Park, PA 16802, USA

*Corresponding authors: mut11@psu.edu, vhc2@psu.edu, phanm@usf.edu

**Dilute magnetic semiconductors, achieved through substitutional doping of spin-polarized transition metals into semiconducting systems, enable experimental modulation of spin dynamics in ways that hold great promise for novel magneto-electric or magneto-optical devices, especially for two-dimensional systems such as transition metal dichalcogenides that accentuate interactions and activate valley degrees of freedom. Existing 2D ferromagnets are a source of fascinating fundamental physical phenomena[1-5], although these may require cryogenic temperatures[1-2], protection from the air[1, 4-5], or a modest external field to induce magnetization[2]. Two-dimensional ferromagnets that are metallic[3] cannot access the rich electronic and optical phenomena open to semiconductors. Practical applications of 2D magnetism will likely require room-temperature operation, air stability, and (for magnetic semiconductors) the ability to achieve optimal doping levels without dopant aggregation. Here we describe room-temperature ferromagnetic order obtained in semiconducting vanadium-doped tungsten disulfide monolayers produced by a reliable single-step film sulfidation method across an exceptionally wide range of vanadium concentrations, up to 12 at% with minimal dopant aggregation. These monolayers develop p-type transport as a function of vanadium incorporation and rapidly reach ambipolarity. Ferromagnetism peaks at an intermediate vanadium concentration of a few atomic percent and decreases for higher concentrations, which is consistent with quenching due to orbital hybridization at closer vanadium-vanadium spacings, as supported by transmission electron microscopy, magnetometry and first-principles calculations. Room-temperature two-dimensional dilute magnetic semiconductors provide a new component to expand the**

**functional scope of van der Waals heterostructures and bring semiconducting magnetic 2D heterostructures them into the realm of practical application.**

Dilute magnetic semiconductors (DMS) have been realized in bulk III-V systems such as Mn-doped GaAs, with long efforts towards achieving room-temperature ferromagnetism in the face of dopant clustering and phase segregation.[6],[7],[8] Pioneering efforts towards achieving DMS behavior in semiconducting transition metal dichalcogenides (TMD)[9],[10] has recently burgeoned on atomically thin samples towards the introduction of magnetic transition metal ions such as V[11], Ni[12], Co[13] and Mn[14] into the host lattice. While first-principles calculations predicts tunable ferromagnetism in V-doped $MoS_2$[15], $WS_2$[16] and $WSe_2$[17] monolayers, reliable experimental realization of ferromagnetism in thin-film samples has been challenging[11]. Intrinsic ferromagnetism has been confirmed in semiconducting monolayer $CrI_3$[1, 4-5] and insulating few-layer $Cr_2Ge_2Te_6$[2] at cryogenic temperatures. Moreover, a transition from paramagnetic to ferromagnetic in vanadium diselenide has been recorded when this metallic material was isolated in monolayer form[3]. Air sensitive monolayer $VSe_2$ has displayed ferromagnetic order even at and above room temperatures[18]. Monolayer samples provide compelling advantages in the characterization of atomic structure and integration into van der Waals heterostructures[10]. Monolayer tungsten disulfide is furthermore a direct-gap semiconductor with high photoluminescence yield that can achieve a reasonably high on/off current ratio ($>10^5$) in field-effect transistor geometries[19]. Reliable substitutional cation doping of $WS_2$ and its sister material $MoS_2$ can induce degenerate n-type (rhenium[20]) and p-type (carbon[21] and niobium[22]) conduction. Beyond simply introducing charge carriers, a judicious choice of dopant may also introduce spin polarization. Scalable and controllable synthesis of single-phase monolayer DMS's with ferromagnetic ordering at room temperature could thus provide a new component for van der Waals heterostructures that express novel modes of magneto-electric and magneto-optical response.[23],[24]

We report the single-step and atmospheric pressure deposition (via film sulfidation) of high-quality V-doped $WS_2$ monolayers exhibiting room-temperature ferromagnetism. Aberration-corrected high-resolution scanning transmission electron microscopy (AC-HRSTEM) and X-ray photoelectron spectroscopy (XPS) reveal substitutional vanadium concentrations up to 12 atomic percent (at%) without substantial structural deformation or degradation. Interestingly, vanadium doping (or alloying) reduces the optical bandgap and induces p-branch transport that reaches ambipolarity. What appears to be intrinsic ferromagnetic order is achieved at room temperature, with a maximum coercivity ($H_c$=130 Oe) and saturation magnetization at an intermediate vanadium concentration (~2 at%). First-principles calculations suggest that magnetism can be further strengthened by optimizing the distribution of dopant-dopant neighbor separations, and also reveal how spin polarized impurity levels breaks the valley degeneracy. These results now establish a promising route to room-temperature 2D spintronic devices.

Pristine and V-doped $WS_2$ monolayers were synthesized by chemical vapor deposition (CVD)[25] (schematics at upper left in Figure 1) with ammonium metatungstate (($NH_4$)$_6$$H_2$$W_{12}$$O_{40}$·x$H_2$O, AMT) and vanadium oxide sulfate (VO[$SO_4$]) acting as W/V cation precursors, in conjunction with sodium cholate ($C_{24}H_{39}NaO_5$ · x$H_2$O) as a surfactant salt, all dissolved in deionized water. VO[$SO_4$] directly supplies $V^{4+}$ ions (i.e. oxovanadium, $VO^{2+}$), which may be the key to reproducibly achieving a wide range of substitutional V concentrations in $WS_2$; results for alternative precursors are described in Supplementary Figure 1. The prepared solutions were spin-coated on oxidized Si substrates, followed by sulfidation at 800°C under atmospheric pressure (Figure 1 and Methods). As-grown pristine and V-doped $WS_2$ monolayers have regular triangular shapes 10 to 50 μm across (Supplementary Figure 2a). The dopant's elemental fingerprints from TEM electron energy loss spectroscopy (EELS) at the vanadium $L_{2,3}$



edges of 513 eV and 521 eV were further confirmed by the observation of a vanadium Kα peak at *circa* 4.95 keV in STEM energy-dispersive X-ray spectroscopy (STEM/EDS, Supplementary Figure 2c), and XPS elemental analyses (Supplementary Figures 2d-f). An oxygen peak is unavoidable in the EELS spectra, due to oxidation of the sample surface and film support. Keeping the overall volume of the precursor solution constant, the vanadium precursor concentration was varied from zero to $1\times10^{-5}$, $1\times10^{-4}$, $1\times10^{-3}$ and $5\times10^{-3}$ mol/L; (higher vanadium concentrations triggered precipitation upon mixing W and V precursor solutions, leading to degradation of the $WS_2$ monolayer's crystallinity, edge regularity, and flatness). XPS analysis measured overall doping levels of approximately ~1.5 at% and ~10 at% for the $1\times10^{-4}$ and $1\times10^{-3}$ mol/L solutions, while the vanadium concentration in the $1\times10^{-5}$ sample was below the XPS detection limit. Direct enumeration of vanadium dopants by atomic resolution TEM (Figure 2 and Supplementary Figures 3a and b) yields more sensitive results, with average vanadium concentrations of 0.4, 2.0, 8.0, and 12.0 at%. obtained for V-doped $WS_2$ grown using the $1\times10^{-5}$, $1\times10^{-4}$, $1\times10^{-3}$ and $5\times10^{-3}$ mol/L solutions, respectively; no detectable vanadium was found in pristine $WS_2$ monolayers grown from 0 mol/L solution. The 0.4, 2.0, and 8.0 at% samples showed large area monolayer morphology and were subject to further detailed optical, electronic, structural, and magnetic characterization as described below. The TEM images of Supplementary Figure 4 show how the vanadium concentration varies from the center to edge of the triangular monolayer flakes, for example from 5.5 at% at the center to 2.1 and 1.7 at% in the middle and outer regions of a flake grown from $1\times10^{-4}$ mol/L solution; the intermediate (middle) value is used to denote each sample, and this is the region from which photoluminescence (PL) Raman spectroscopy and electrical transport measurements are generally taken.



Representative Raman spectra of pristine and V-doped WS$_2$ monolayers were obtained using excitation wavelengths of 532 nm (Figure 1) and 488 nm (Supplementary Figure 5). Pristine WS$_2$ monolayers exhibit the representative E′(Γ) and A$_1$′(Γ) first-order phonon modes at 355 and 417 cm$^{-1}$, respectively.[21] Both E′(Γ) and A$_1$′(Γ) blueshift as a function of vanadium concentration, which is consistent with previously reported spectra of V-doped MoS$_2$.[26] In V-doped WS$_2$ samples, the defect-activated longitudinal acoustic mode (LA(M)) gradually emerges as the vanadium concentration increases, indicating lattice disorder induced by V dopants.[27] A high-intensity 2LA(M) second-order double resonance peak involving two longitudinal acoustic phonons[28] characteristic for pristine WS$_2$ monolayers was progressively suppressed upon increasing vanadium concentration, indicating substantial changes in the WS$_2$ electronic structure which drive the system out of resonance.[21],[28] Pristine monolayers of WS$_2$ show an intense PL peak at 1.97 eV corresponding to the A exciton[29]. The optical gap decreases under increasing vanadium doping, while the PL peak broadens (likely due to lattice disorder from dopants possibly accompanied by vacancies), and drops in intensity. This evolution of the PL response is consistent with the Raman results discussed above.

Figure 1 shows the drain current (I$_{DS}$) versus back-gate voltage (V$_{BG}$) transport characteristics for 0, 0.4, 2, and 8 at% vanadium monolayers at various drain voltages (V$_D$), measured by back-gated FET geometry (detailed in Methods). The 0 at% device shows unipolar electron conduction with no hole current, as expected for a pristine WS$_2$-based FET. A small hole branch emerges at 0.4 at% doping, with increasing hole currents at higher doping levels as the threshold voltage shifts from 2.9 volts in the pristine system to 5.6, 6.3 and 11.3 volts in devices with progressively higher doping levels (we extract V$_{TN}$ by the iso-current method for a current of 1 nA/µm at V$_D$=1 volt). The systematic threshold shift confirms p-type doping, with



the most heavily doped sample demonstrating close-to-symmetric ambipolar transport. Thermal boundary conductance ($h_K$) measured at the Al / V-WS$_2$ / SiO$_2$ interface using time-domain thermo-reflectance (Materials and Methods, Supplementary Figure 6) reveals a significant improvement in heat dissipation under vanadium doping that could be helpful in device applications.

High-angle annular dark field (HAADF)-STEM imaging confirms the presence of substitutional V atoms at W sites (written $V_W$) in WS$_2$, and reveals surprisingly little dopant aggregation. Under Z-contrast imaging vanadium is easily distinguished from much heavier tungsten, and its concentration can be extracted by statistical analysis of HAADF-STEM images (Figure 2 and Supplementary Figure 7) at multiple locations on each flake. Dopant aggregation is modest even at 12 at% vanadium. Occasional aggregations into stripes at high doping levels (Supplementary Figure 3), tend to align to the proximate outer edge of the flake, thus suggesting that edge energetics/kinetics influence dopant aggregation[30]. Comparing experimental STEM images to image simulations (simulation details in Materials and Methods, Figure 2 and Supplementary Figure 7), clearly reveals elemental identities by contrast differences (W > 2S > V > 1S). At vanadium concentrations of 8 at% or above (Supplementary Figure 3e) sulfur vacancies are more likely to be coupled to V atoms (written $V_W+S_{vac}$), which is consistent with prior work on TMD alloys [11], [31] and first-principles calculations described below.

The magnetic properties of pristine and V-doped WS$_2$ monolayer samples were measured by a vibrating sample magnetometer. To exclude unwanted effects on the magnetization versus magnetic field (M-H) loops that can arise from subtracting diamagnetic and paramagnetic backgrounds[32],[33], Figure 2 presents as-measured M-H loops at 300K and deduces the saturation magnetization and coercive field ($H_C$) directly from these loops. The pristine WS$_2$



sample exhibits a very weak ferromagnetic signal, which we tentatively ascribe to undercoordinated sulfur atoms at crystallographic defects (e.g. edges)[34], on a diamagnetic background. Vanadium doping of 0.4 at% greatly increases the ferromagnetic signal ($M_S$ and $H_C$) with further strengthening at 2 at% doping and then a much weaker ferromagnetic response at 8 at% vanadium. The ferromagnetism observed in V-doped samples is too strong to originate from edge effects, and its dependence on vanadium concentration suggests an origin in local moments associated with unpaired electrons in vanadium d orbitals. [11],[16],[32] The 2 at% $WS_2$ sample shows large, clear hysteresis loops at all temperatures from which $M_S$ and $H_C$ are extracted and plotted as a function of temperature (the raw loops are close to square when rotated to account for the diamagnetic background). The saturation magnetization and coercivity both increase with decreasing temperature, with an interesting non-monotonicity of both around 150–200K.

Density functional theory calculations show local moments on substitutional vanadium atoms whose spin polarization and coupling are sensitive to the relative placement of dopants, a behavior similar to that seen in other computational investigations of doped TMDs[15],[16],[35]. A single vanadium dopant in a 7×7 supercell hosts a local moment of 0.67$\mu_B$ with vanadium $d_{z^2}$ character that is associated with a partially occupied spin-split defect level sitting ~74 meV below the valence band maximum, as shown in Figure 3. Table 1 compiles the results of the interaction between two vanadium dopants in this supercell. At nearest and next-nearest neighbor separations the vanadium defect states hybridize strongly and local moments are quenched. At larger separations, a clear preference for a ferromagnetic alignment emerges. Considering that ferromagnetic order in the experimental system will benefit from additional neighbor interactions beyond the single pair partner in the current model, the ~10 meV stabilization for parallel alignment over anti-parallel is reasonably consistent with the observed room-temperature



ferromagnetic order. The origin of the non-monotonicity in $M_S$ and $H_C$ around 150–200K, where the saturation magnetization drops just as the coercivity more rapidly increases, is less clear. Speculatively, it might reflect domains of competing anti-ferromagnetic order that reduce $M_S$ while increasing $H_C$ through exchange pinning. The roles of the detailed microscopic distribution of dopants and possible variations in the Fermi level beyond that induced by vanadium alone provide further caveats on the correspondence of computation to experiment. In this regard, the appearance of similar quenching phenomena even in calculations without spin-orbit coupling (Supplementary Table 1, Supplementary Figure 9) lends substantial confidence to the quenching mechanism.

Spin-momentum locking at the valence band top of $WS_2$ interacts closely with the local moments of the vanadium dopants, yielding band crossings or avoided crossings respectively if the spin of a dopant band and the valence band extremum has the same or opposite direction, with the circumstance switching between these two cases as one proceeds from K to –K, which is visible in Supplementary Figure 8. For a single vanadium, the two types of spin-polarized band edges shift in opposite directions in energy such that the valence band maximum (VBM) has an energy difference of 12.6 meV between valleys while the original split in conduction band minima (CBM) collapses in one valley become nearly spin-degenerate. Increasing the doping level to two vanadium atoms in the supercell (with ferromagnetic order), the VBM difference increases to 19.5 meV and the CBM shows uniform spin direction. These band edge offsets arising from the ferromagnetic polarization correspond to an effective external magnetic field on the order of ~100 T.[36]

A statistical analysis of the expected fraction of magnetically polarized dopants in a random alloy (Supplementary text 9) suggests that the optimal doping level to obtain maximum



saturation magnetization is intermediate between the 2at% and 8at% samples examined experimentally. First-principles calculations indicate that the nearest-neighbor separation of a pair of vanadium dopants is 50–90 meV more stable than larger separations, meaning a modest energetic preference for dopant aggregation as is evidenced experimentally by occasional stripes at higher vanadium concentrations. This suggests that the choice of precursor may be an important factor in atomic scale dopant structure optimization (Supplementary Figure 1). The rare occurrence of V aggregation in moderately doped samples suggest that synthesis is far from equilibrium, presumably a key to incorporating V dopants into the $WS_2$ host in the first place. Finally, the observed spatial correlation between sulfur vacancies and vanadium dopants (Supplementary Figure 3) is consistent with computational results that sulfur vacancies bind to a single vanadium by ~0.04 eV and to a nearest-neighbor V-V dimer by ~0.71 eV.

This study successfully develops a universal, scalable and controllable synthesis route for V-doped $WS_2$ atomic layers as a dilute magnetic semiconductor, with intrinsic ferromagnetic ordering at room temperature. As the vanadium concentration increases, V-doped $WS_2$ monolayers exhibit a reduction of the optical bandgap and the emergence of p-type transport, reaching ambipolarity. The vanadium doping induces inherent ferromagnetic ordering at room temperature, with the strongest ferromagnetic signal for the moderately doped (2 at%.) sample. The non-monotonicity of the magnetization as a function of doping level dopant level is explained by a combination of atomic resolution TEM imaging and DFT calculations which show how hybridization between dopant defect states quenches the magnetic moment. An effective Zeeman shift corresponding to ~100 T is observed in the calculated bandstructure. Such dilute magnetic semiconductors based on magnetic-element-doped transition metal dichalcogenides exhibit great promise as future spintronic/valleytronic devices, with novel



magneto-electric and magneto-optical responses. Furthermore, they constitute a new set of atomically thin layers that could be integrated in multifunctional van der Waals heterostructures.

Methods and Experimental Section

*Synthesis of pristine and vanadium doped $WS_2$ monolayers:* 0.05 g ammonium metatungstate (($NH_4$)$_6$$H_2$$W_{12}$$O_{40}$ · $xH_2O$, AMT) and 0.2 g sodium cholate ($C_{24}H_{39}NaO_5$ · $xH_2O$) powders were dissolved in 10 ml water to form a tungsten precursor solution. 0.05g vanadyl sulfate (VO[$SO_4$]) powder were dissolved in 10mL deionized water to form vanadium precursor ($1 \times 10^{-2}$ mol/L). The W and V with different concentrations were controlled to form solution-based cation precursors. We drop casted the precursor solution onto a $SiO_2$/Si substrate followed by spin-coating for 1 min with 3000 rpm. The film sulfidation process was carried out at atmospheric pressure in a quartz reaction tube (1" inner diameter) with sulfur powder (400 mg) heated upstream at low temperature (220 °C, heated up using a heating tape), and the cation precursor, spin-coated on the $SiO_2$/Si substrates, at the high temperature (825 °C) zone. Ultrahigh purity argon was employed as the carrier gas. The furnace was then allowed to cool to room temperature naturally after 15 min synthesis.

*Materials Characterization*: A Renishaw inVia microscope with a Coherent Innova 70C argon-krypton laser at the excitation of 488 nm and a LabRAM HR evolution (Horiba) equipped with a 532 nm laser were used for acquiring the Raman and photoluminescence (PL) spectra using a backscattering configuration and an 1800 line/mm grating. X-ray Diffraction (XRD) was taken with PANalytical Empryean X-Ray Diffractometer with a Cu source. X-ray photoelectron spectroscopy (XPS) experiments were performed using a Physical Electronics VersaProbe II instrument. The binding energy axis was calibrated using sputter cleaned Cu foil (Cu $2p_{3/2}$ = 932.7 eV, Cu $2p_{3/2}$ = 75.1 eV). UV-Vis absorption spectra were transformed from reflectance measurements, which were acquired on Perkin-Elmer Lambda 950 with a universal reflectance accessory (URA). STEM-EDS of the samples was performed in a FEI Talos F200X microscope with a SuperX EDS detector, operating at 200kV. Aberration-corrected STEM imaging and EEL spectroscopy were performed using a FEI Titan $G^2$ 60-300 microscope, operated at 80kV with double spherical aberration correction, offering the sub-angstrom imaging resolution. A HAADF detector with a collection angle of 42-244 mrad, camera length of 115 mm, beam current of 45pA and beam convergence of 30 mrad were used for STEM image acquisition. For the HAADF-STEM images, Gaussian blur filter (r=2.00) was applied (by the ImageJ program) to eliminate noise and enhance the visibility of structural details, while the line profiles of ADF intensity were captured by analyzing raw STEM images. Atomic resolution STEM image simulations were conducted by using the QSTEM package[37], the case of vanadium dopant coupled with sulfur monovacancy is set to be three symmetric neighboring sulfur monovacancies in the crystal structure to simplify the simulation process. The applied parameters, acceleration voltage, convergence angle and inner/outer angle for the HAADF detector and spherical aberration ($C_3$ and $C_5$), were all adjusted according to the experimental conditions.

*DFT calculations:* Spin-orbit-coupled DFT calculations were implemented in the Vienna Ab-initio Simulation Package(VASP).[38],[39],[40] A 7×7 supercell of $WS_2$ was tested with different V doping levels. The z-axis cell dimension was 15 Å to isolate a layer from its in periodic images. The exchange-correlation was treated under GGA PBE approximation[41] with PAW method[42]. The energy cutoff in all calculation was 700 eV and the k-point sampling was set as 4×4×1 centered at Γ. The $WS_2$ unit cell lattice constant calculated as 3.188 Å matches with previous work[43] and was fixed for doped $WS_2$ since the doping level is not high enough to significantly change the lattice constant. The residual force after relaxation was smaller than 0.01 eV/ Å for all atoms. All visualizations were done with VESTA[44] and pymatgen[45].

As the experimental distribution of vanadium dopants is irregular and covers a wide range of pairwise separations, we did not model the system with a regular array of dopants at uniform mutual separations but instead examined a dopant pair hosted within a large 7 ×7 supercell across a range of separations, so that we can elucidate their interactions on a pairwise basis. The two dopants within this supercell are closer to each other than to any periodic replicas. We considered both ferromagnetic (parallel) and antiferromagnetic (anti-parallel) initial spin configurations for two vanadium dopants in a supercell, with the system also being able to converge self-consistently into an unpolarized state in the case of moment quenching.

*Electronic Device fabrication:* Pristine and V-doped $WS_2$ triangles were transferred from the growth substrate ($Si/SiO_2$) to a 50 nm thick and atomic layer deposition grown $Al_2O_3$ substrate with Pt/TiN/$p^{++}$Si as the back-gate electrode. All FETs were fabricated with a channel length of 1 μm with 40 nm Ni/30 nm Au as the source/drain contact electrodes defined using standard electron-beam lithography process.

*Thermal transport measurements:* To examine the thermal boundary conductances ($h_K$) of devices contingent on the use of doped $WS_2$, we deposited a nominally 80 nm Al film via electron beam evaporation. We measured the total conductance of the Al/doped $WS_2$/$SiO_2$ interface via time-domain thermoreflectance (TDTR). The specific analyses can be found elsewhere.[46] In our implementation, the 808.5 nm output of a Ti:Sapphire oscillator is spectrally separated into high-energy pump and low-energy probe paths. The pump is electro-optically modulated at 8.4 MHz, and creates a frequency dependent heating event at the sample surface.

The probe is mechanically delayed in time, and monitors the change in reflectivity due to the pump-induced heating event (i.e., thermoreflectivity) as a function of delay time. Both the pump and probe are concentrically focused through a 10x objective, yielding $1/e^2$ diameters of 14 and 11 μm, respectively. The data are fit to the radially symmetric heat diffusion equation to extract the conductances at the Al/doped $WS_2$/$SiO_2$ interface.

*Magnetic measurements:* Temperature- and magnetic field-dependent magnetization measurements were carried out in a Physical Property Measurement System (PPMS) from Quantum Design with a vibrating sample magnetometer (VSM) magnetometer over a temperature range of 2 – 350 K and fields up to 9 T.

Supporting Information

Supporting Information is available from the Wiley Online Library.

Acknowledgments

This work was mainly supported by the Air Force Office of Scientific Research (AFOSR) through grant No. FA9550-18-1-0072 and the NSF-IUCRC Center for Atomically Thin Multifunctional Coatings (ATOMIC). Theory efforts were supported by the Two-Dimensional Crystal Consortium (2DCC-MIP), a National Science Foundation Materials Innovation Platform, through award DMR-1539916. M.H.P. acknowledges support from the U.S. Department of Energy, Office of Basic Energy Sciences, Division of Materials Sciences and Engineering under Award No. DE-FG02-07ER46438 and the VISCOSTONE USA under Award No. 1253113200. The authors are grateful to Jeff Shallenberger from Material Characterization Laboratory at Penn

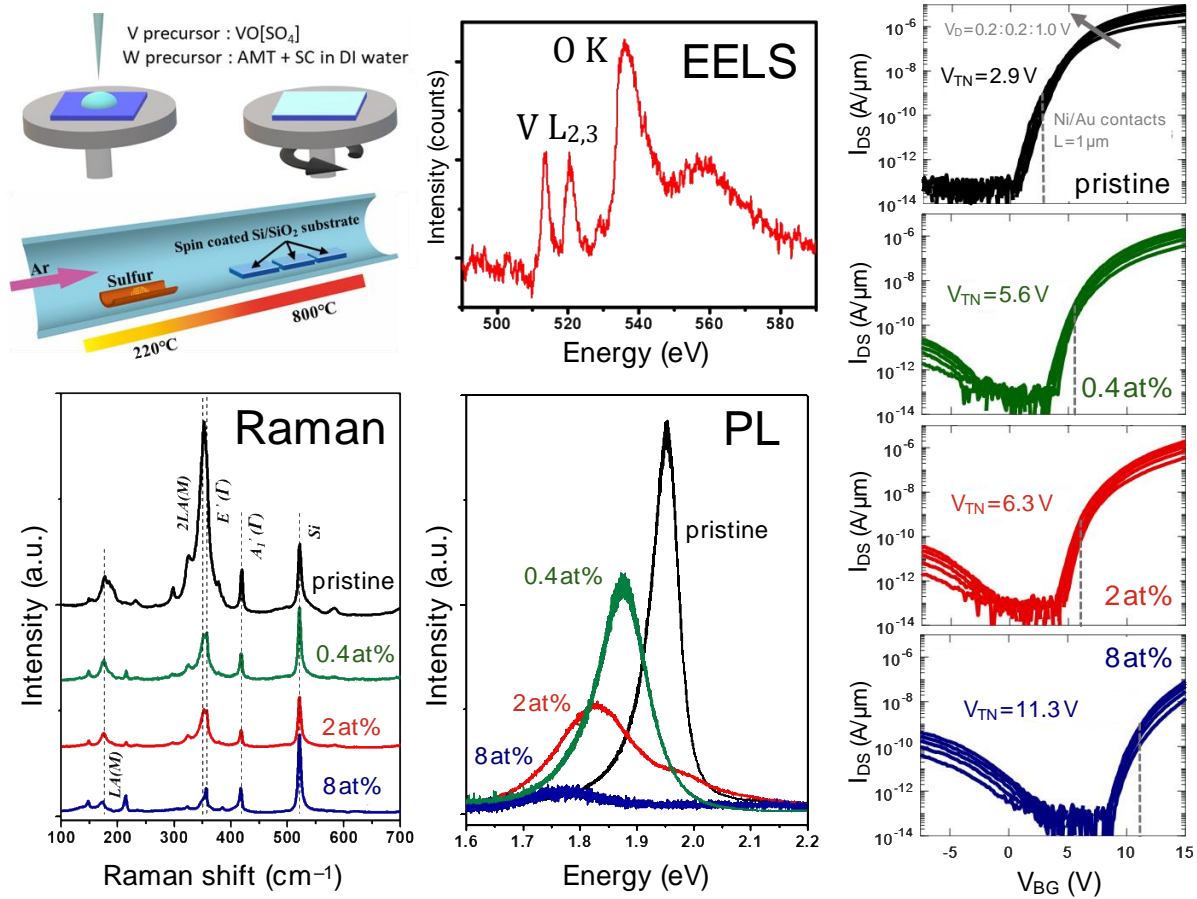

Figure 1| One-step synthesis of monolayer V-doped $WS_2$, optical and electronic properties, as described schematically at upper left, yields a TEM/EELS spectrum with a prominent vanadium $L_{2,3}$ edge. A loss of double resonance in Raman (under 532nm excitation) and pronounced change in photoluminescence response reflect a change of electronic structure as a function of V doping. Back-gated V-doped $WS_2$ field effect transistors were fabricated on a 50nm thick $Al_2O_3$ substrate with a $Pt/TiN/p^{++}$ back-gate electrode for each doping level. Drain current ($I_{DS}$) versus back-gate voltage ($V_{BG}$) (obtained for drain voltages from 0.2 to 1V in 0.2V steps) show a steady shift in threshold voltage across different doping levels and achieve close-to-symmetric ambipolar conduction in heavily doped $WS_2$.

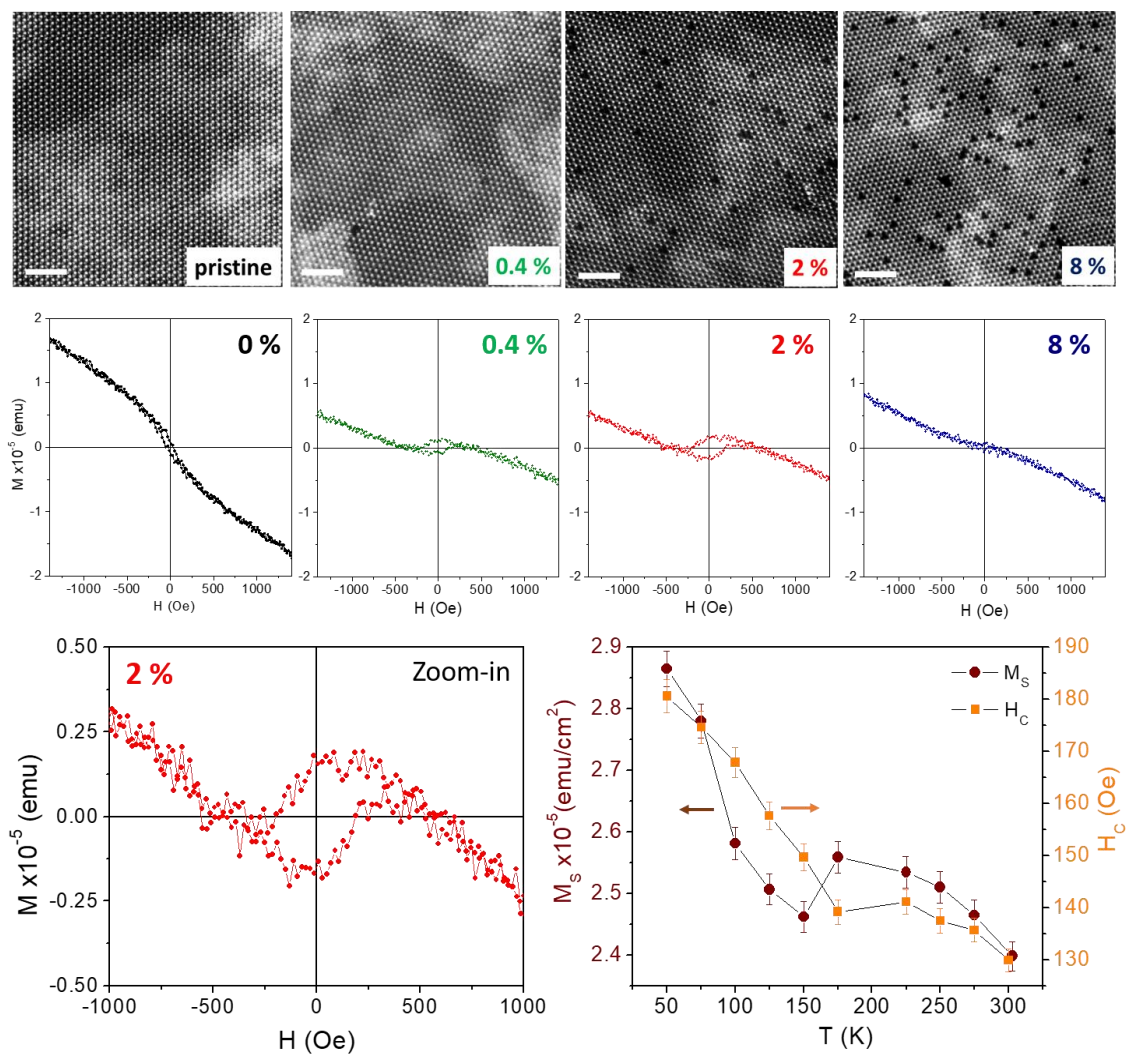

**Figure 2|** **Atomic resolution HAADF-STEM images (upper) and magnetization versus field loops (middle)** at 300 K for of pristine and vanadium-doped $WS_2$ monolayers at 0.4, 2, and 8 at% vanadium, scale bars are 2nm. **Bottom**: an expanded view of the hysteresis loop for the 2at% sample and its temperature-dependent saturation magnetization ($M_S$) and coercivity ($H_C$).

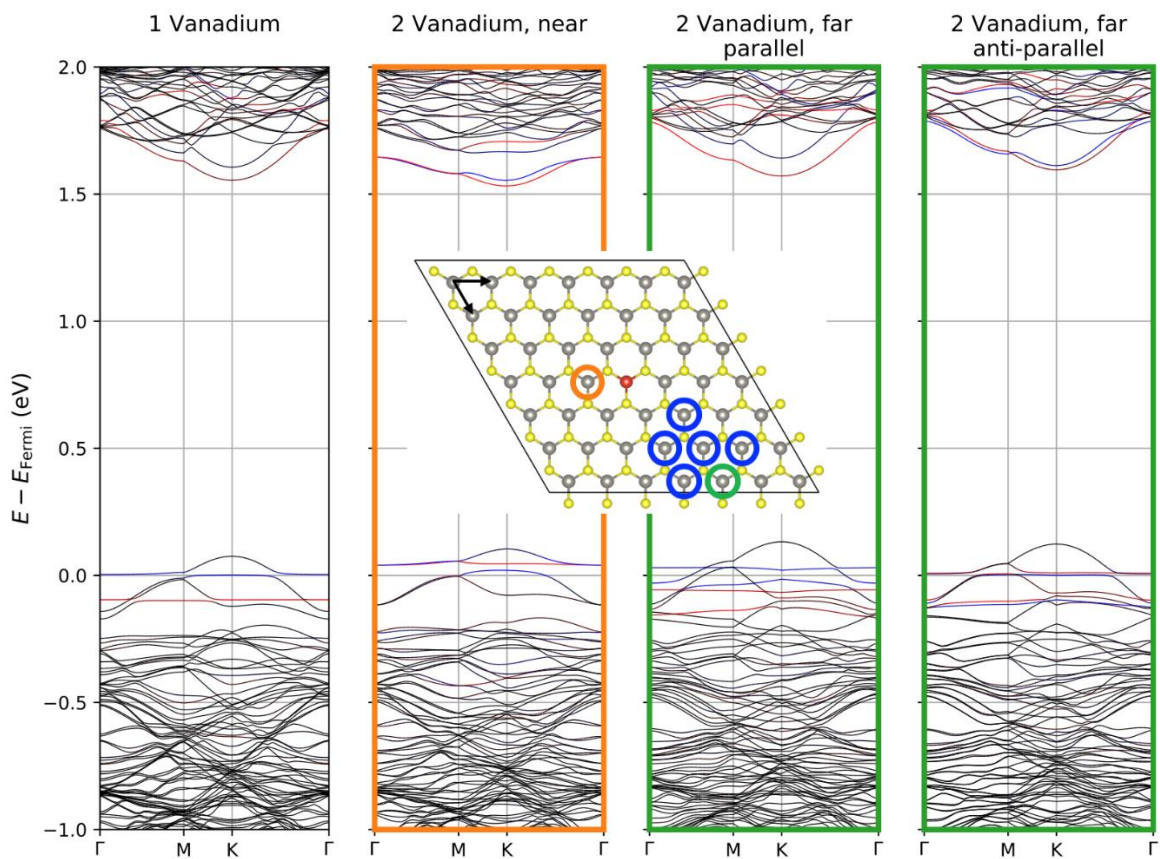

Figure 3| DFT calculation results for V-doped $WS_2$ monolayers. All non-equivalent positions for the second V dopant are circled. The bandstructures of a single vanadium and two vanadium atoms with the nearest and the farthest separations are plotted. Other, symmetry non-equivalent k-directions (due to low-symmetry dopants placements) look similar (Supplementary material Figure S7) Red/blue indicate spin up/down polarization for states of V $d_{z^2}$ character. The arrows in the supercell show the primitive-cell lattice vectors used to label dopant pairs in Table 1.

Table 1| Net moments and energies for vanadium dopant pairs. Dopant pairs are labeled by their separation in lattice coordinates and colored in reference to the supercell in Figure 3. Systems were initialized with either parallel or anti-parallel local moments around the two dopants. Moments after self-consistent iterations are perpendicular to the plane except for the (0,2) separation, which is 76° away from this axis. For the closest and next-closest dopant separations (★), the lowest energy state examined has no spatially resolvable spin texture. "--" means that both parallel and anti-parallel initial spin textures converge to the same self-consistent state.

| Dopant pair in lattice coordinatess | Pair separation (Å) | Energy of the most stable spin texture (meV) | Net magnetic moment ($\mu_B$) | Energy of competing spin texture (meV) | Moment of competing spin texture ($\mu_B$) |
|---|---|---|---|---|---|
| –1, 0 | 3.19 | 0  (★) | 0.00 | -- | -- |
| 1, 1 | 5.52 | 63.9 (★) | 0.00 | 67.5 (↑↑) | 0.14 |
| 0, 2 | 6.38 | 48.2 (↑↑) | 0.93 | -- | -- |
| 1, 2 | 8.44 | 71.1 (↑↑) | 1.18 | 84.4 (↑↓) | 0.00 |
| 0, 3 | 9.57 | 86.8 (↑↑) | 1.22 | 95.4 (↑↓) | 0.03 |
| 2, 2 | 11.02 | 88.8 (↑↑) | 1.24 | 93.6 (↑↓) | 0.07 |
| 1, 3 | 11.49 | 86.5 (↑↑) | 1.23 | 93.9 (↑↓) | 0.11 |

*Table of Contents*

Room-temperature ferromagnetic order was obtained in semiconducting vanadium-doped tungsten disulfide monolayers. A reproducible and atmospheric pressure film sulfidation growth method led to doping concentration tunability in samples that preserve their properties while been stored in air. These monolayers developed p-type transport as a function of vanadium incorporation and rapidly reach ambipolarity. The ferromagnetic behavior in this dilute semiconductor was modeled and understood through first-principles calculations.

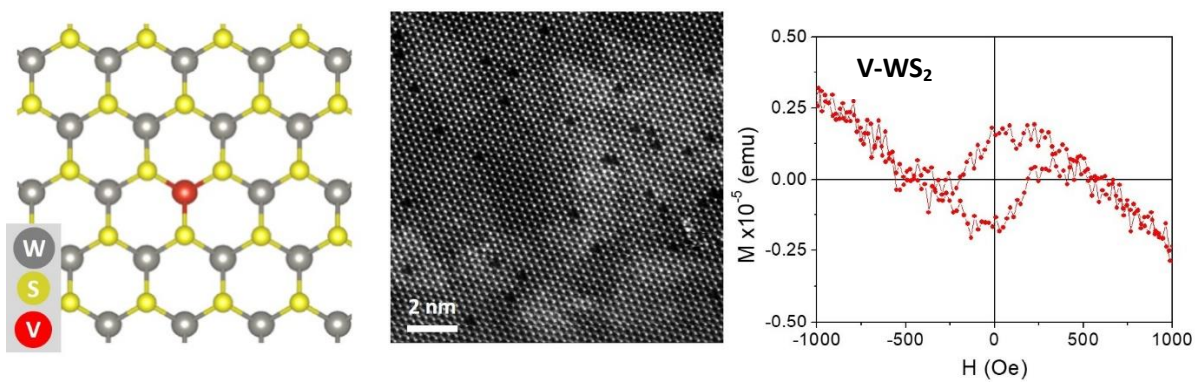

# Monolayer Vanadium-doped Tungsten Disulfide: A Room-Temperature Dilute Magnetic Semiconductor

Fu Zhang[1,2], Boyang Zheng[3], Amritanand Sebastian[4], Hans Olson[5], Mingzu Liu[2,3], Kazunori Fujisawa[2,3,6], Yen Thi Hai Pham[7], Valery Ortiz Jimenez[7], Vijaysankar Kalappattil[7], Leixin Miao[1], Tianyi Zhang[1], Rahul Pendurthi[4], Yu Lei[1,2], Ana Laura Elías[2,3], Yuanxi Wang[3,8], Nasim Alem[1], Patrick E. Hopkins[5], Saptarshi Das[1,4], Vincent H. Crespi[3,8*], Manh-Huong Phan[7*], Mauricio Terrones[1,2,3,9*]

[1]Department of Materials Science and Engineering, The Pennsylvania State University, University Park, PA 16802, USA
[2]Center for 2- Dimensional and Layered Materials, The Pennsylvania State University, University Park, PA 16802, USA
[3]Department of Physics, The Pennsylvania State University, University Park, PA 16802 USA
[4]Engineering Science and Mechanics, The Pennsylvania State University, University Park, PA 16802, USA
[5]Department of Mechanical and Aerospace Engineering, University of Virginia, Charlottesville, VA 22904, USA
[6]Research Initiative for Supra-Materials, Shinshu University, 4-17-1 Wakasato, Nagano 380-8553, Japan
[7]Department of Physics, University of South Florida, Tampa, Florida 33620, USA
[8]2D Crystal Consortium, The Pennsylvania State University, University Park, PA 16802, USA
[9]Department of Chemistry, The Pennsylvania State University, University Park, PA 16802, USA

*Corresponding authors: mut11@psu.edu, vhc2@psu.edu, phanm@usf.edu

This file includes:

Supplementary Figures 1-9

Supplementary Table 1

*Supplementary Materials*

1. Discussion of several V doping precursors for V-doped $WS_2$ monolayers synthesis.

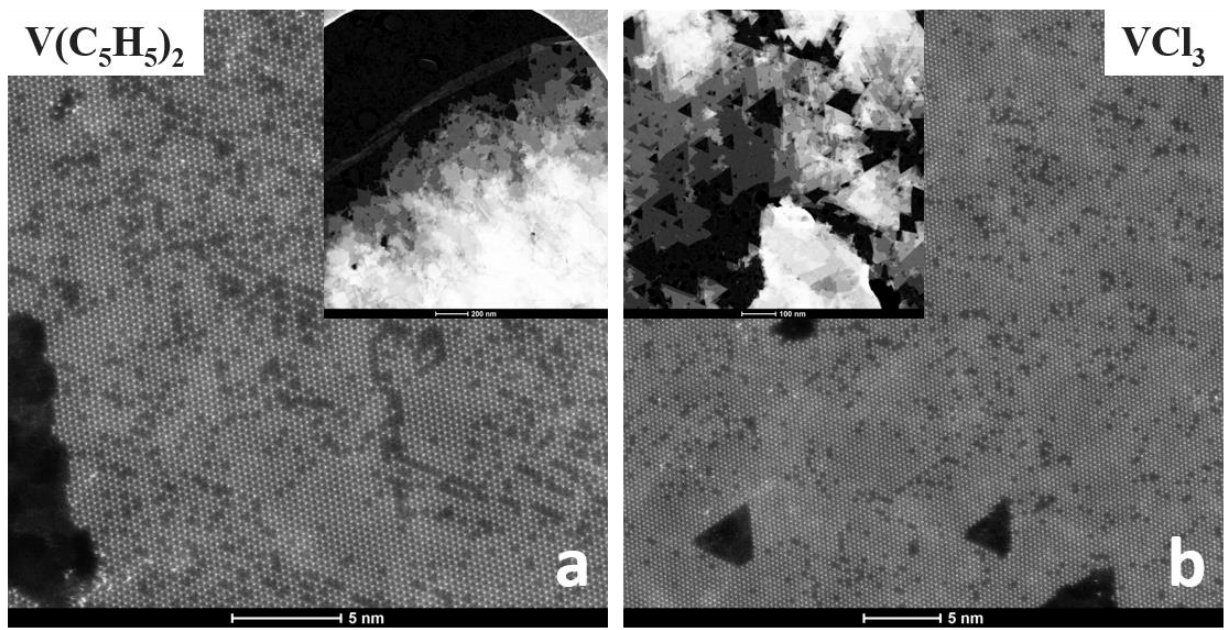

Figure 1 HAADF-STEM image of V-doped $WS_2$ by different V precursors: (a) $V(C_5H_5)_2$ and (b) $VCl_3$.

We have studied several vanadium precursors for doping atoms into $WS_2$ monolayers,

1. For vanadocene (II) ($V(C_5H_5)_2$), we realize doping by powder vaporization growth;[1] it could be doped into the lattice with high concentration, however there are carbon contaminations or even a defective graphene layers underlying as-grown TMDs, as reported for other metal-organic precursors.[2] The morphology is degraded, meaning it fails to form large-area monolayer triangles. Atomic-resolution HAADF-STEM images of highly doped materials show extensive stripes of V dopants in the lattice (Supplementary Fig. 1a), which substantially affects the physical properties.

2. For vanadium (III) chloride ($VCl_3$), we realize doping by powder vaporization growth;[1] as this precursor is very air-sensitive, it is handled inside a glove box. Doping could also be realized at high concentration, but the final morphology shows as few-layer $WS_2$ filled with etched holes probably related to the chloride (Supplementary Fig. 1b).

3. Vanadium (IV) oxide sulfate (VO[$SO_4$]) is our primary current focus, with different concentrations of the precursor we can dope into TMDs as high as 12% of V with good morphology control.

4. Vanadium (V) oxide ($V_2O_5$) does not work due high stability and extremely low vapor pressure which impedes sulfurization, at least for sulfur vaporization.

5. Ammonium metavanadate (V) ($NH_4VO_3$) can be dissolved in DI water for spin coating; we have not tried this precursor, but it has been reported by a recent study,[3] which claimed high-concentration doping is possible, yet high-quality large-area monolayer triangles were not found for higher doping concentration.

The distribution of the transition metal dopants in the host TMD lattice is also highly dependent on the synthesis process, as the kinetically driven CVD process may result in segregation and stripe formation of V dopants. Vanadium segregation and striping were consistently detected when using vanadocene (V($C_5H_5$)$_2$) and vanadium (III) chloride ($VCl_3$) as precursors, which would significantly affect the magnetic properties of the materials as revealed by DFT calculations.

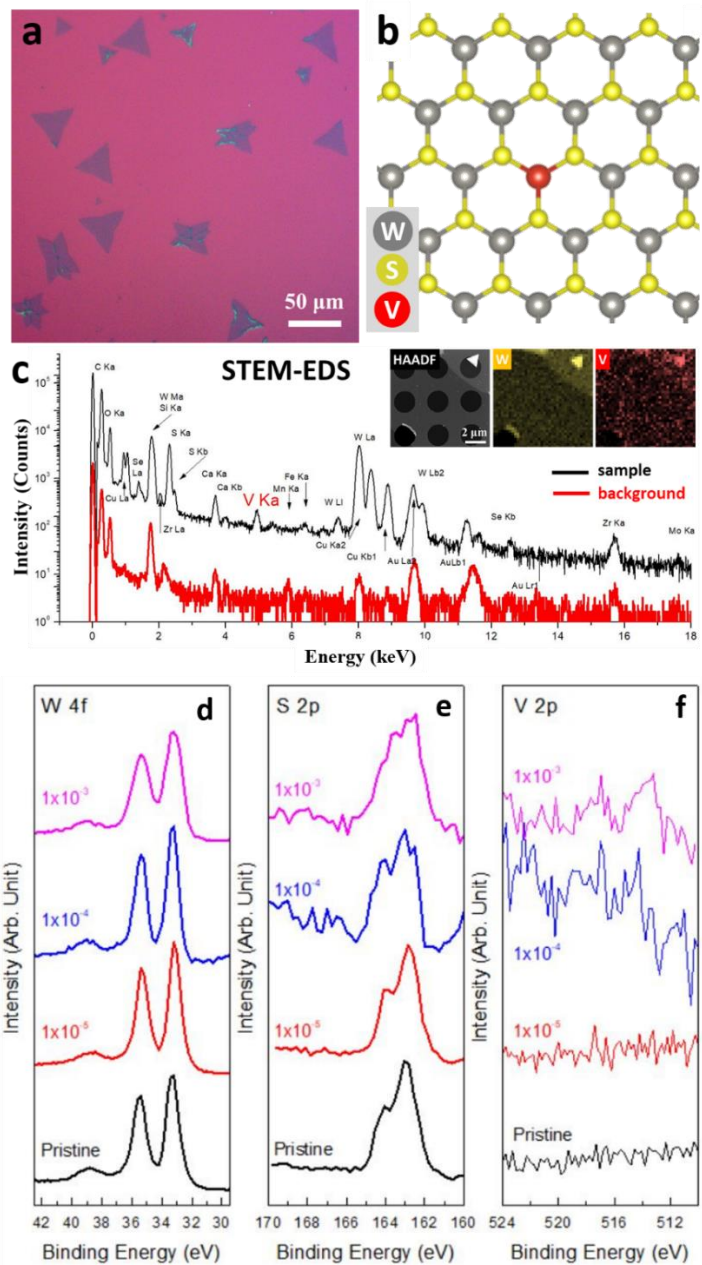

Figure 2 Material characterization of V-doped $WS_2$. (a) Optical microscope image of triangular V-doped $WS_2$ monolayers; (b) Structure schematic of one V atom substitutional doping in $WS_2$ hexagonal lattice; (c) STEM/EDS spectrum of the monolayer V-doped $WS_2$, inset showing the STEM/EDS elemental mappings of the materials. X-ray photoelectron spectroscopy (XPS) elemental analyses of pristine $WS_2$ and V-doped $WS_2$, (d) W 4f, (e) C 1s and (f) V 2p core levels. The vanadium doping levels was below the XPS detection limit for $1\times10^{-5}$ mol/L (V precursor concentration) sample, and approximately 1.5 at% and 10 at% for the $1\times10^{-4}$ and $1\times10^{-3}$ mol/L samples.

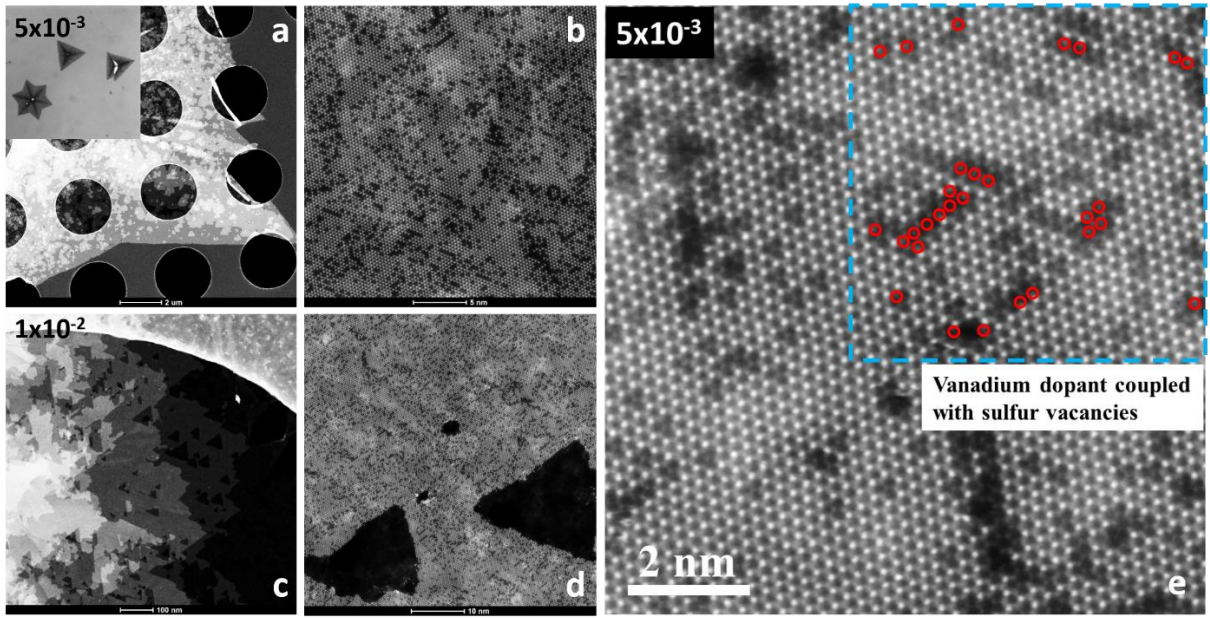

Figure 3 HAADF-STEM images of higher V-doped WS$_2$ monolayers at V precursor concentration of (a) and (b) 5×10$^{-3}$ mol/L (doping level around 12 %) and (c) and (d) 1×10$^{-2}$ mol/L, respectively; (e) indexing of sulfur monovacancies by red circles, it is clear that the sulfur monovacancies tends to be coupled with V substitute dopants and even double V substitute dopants.

4. A vanadium dopant concentration gradient was observed by TEM from this single-step synthesis route, local TEM images in Supplementary Fig. 4 show concentration gradients from the edge, middle to the center of the triangles, for V-WS$_2$ grown using the 1×10$^{-5}$, 1×10$^{-4}$ and 1×10$^{-3}$ mol/L solutions. As analyzed from TEM images, the concentration (by 1×10$^{-4}$ mol/L solution, medium doped) from center, middle to outside area is 5.5 at%, 2.1 at% and 1.7 at%, respectively. The average vanadium concentration used in the main text refers to the middle area local vanadium doping concentration. The corresponding PL and Raman spectroscopy and electrical transport measurements are also conducted in the middle area of the triangles.

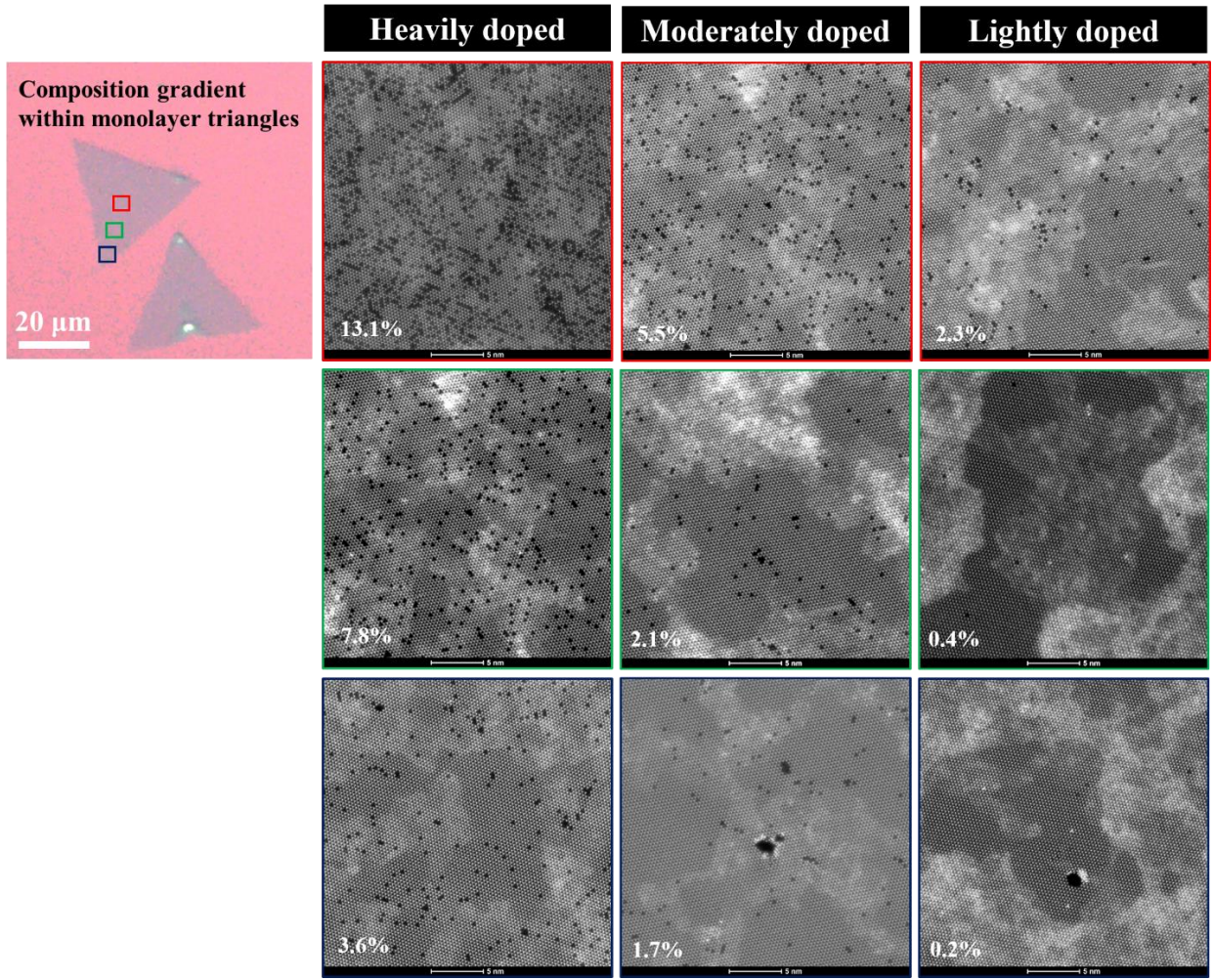

Figure 4 Doping concentration gradients within V-doped monolayer WS$_2$ triangles, local vanadium doping concentrations extracted from HAADF-STEM images at regions of center (red), middle (green) and edge (blue) of the triangles, respectively.

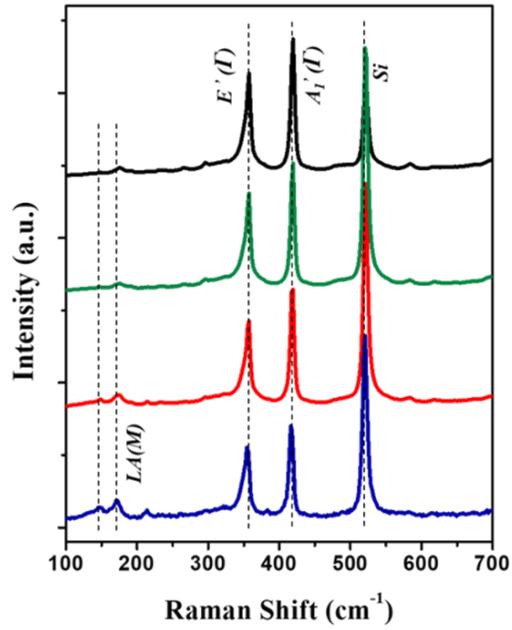

| Peak position (cm$^{-1}$) @488 nm | E' (cm$^{-1}$) | A$_1$' (cm$^{-1}$) |
|---|---|---|
| pristine | 357.6 | 419.3 |
| Lightly doped | 357.3 | 418.9 |
| Moderate doped | 356.2 | 417.7 |
| Heavily doped | 353.4 | 416.0 |
| Peak position (cm$^{-1}$) @532 nm | E'+2LA(M) (cm$^{-1}$) | A$_1$' (cm$^{-1}$) |
| pristine | 352.8 | 419.6 |
| Lightly doped | 352.1 | 418.9 |
| Moderate doped | 351.9 | 418.6 |
| Heavily doped | 356.2 | 417.5 |

Figure 5 Raman spectra of pristine and V-doped WS$_2$ monolayers at 488 nm excitation laser. Table of E'(Γ) and A$_1$'(Γ) peak positions in Raman spectra.

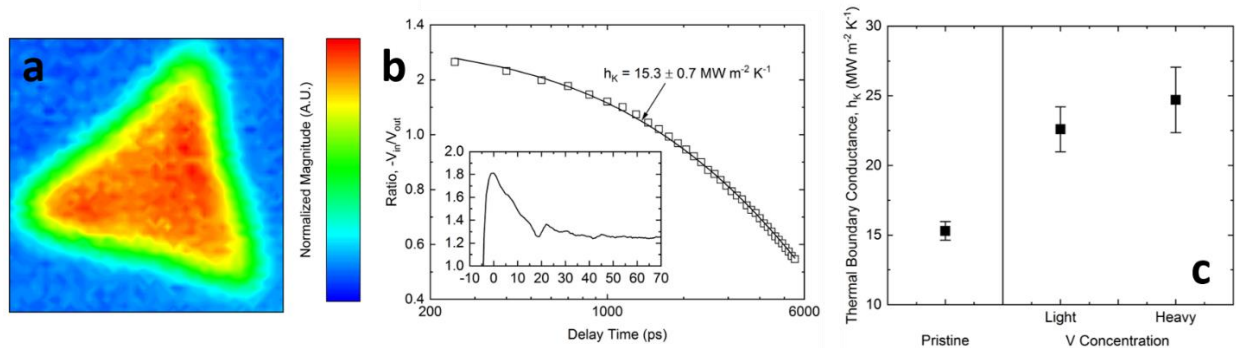

Figure 6 (a) Time-domain thermoreflectance (TDTR) magnitude micrograph of a pristine single-crystal $WS_2$ flake. (b) TDTR model and best fit for the conductance at the Al/pristine $WS_2$/$SiO_2$ interface. The inset shows the picosecond acoustics response at earlier time delays. (c) Results for the thermal boundary conductance at Al/doped $WS_2$/$SiO_2$ interfaces.

To examine the thermal boundary conductances ($h_K$) of devices contingent on the pristine and V-doped $WS_2$ monolayers, we measured the total conductance of the Al/doped $WS_2$/$SiO_2$ interface via time-domain thermoreflectance (TDTR) (detailed in Methods). An example of this magnitude micrograph can be seen in Supplementary Fig. 6a for a pristine $WS_2$ flake. The uniformity of the TR magnitude of the $WS_2$ flake suggests that the conductance is uniform. Full time delay TDTR measurements near the center of the triangles were examined for pristine $WS_2$ and V-doped $WS_2$ samples. The TDTR curve and best fit are shown in Supplementary Fig. 6b for the pristine flake, where the inset shows the short delay time picosecond acoustic response,[7] with which we used to extract the thickness of our Al transducer. This type of measurement was performed on lightly- (0.4%) and heavily-doped (8%) V-$WS_2$ crystals as well. The final results are shown in Supplementary Fig. 6c for all flakes measured, where an increase in thermal conductance of the Al/$WS_2$/$SiO_2$ interface was observed as the V doping concentration increased. The enhancement in conductance is correlated with the V substitutional sites in the $WS_2$ lattice. It is assumed that the inclusion of V dopants alters the local phonon density of states, allowing for an improved thermal conductance as the concentration of V is increased. Monolayer V-doped $WS_2$ with improved heat dissipation is promising as an electronic circuit component.

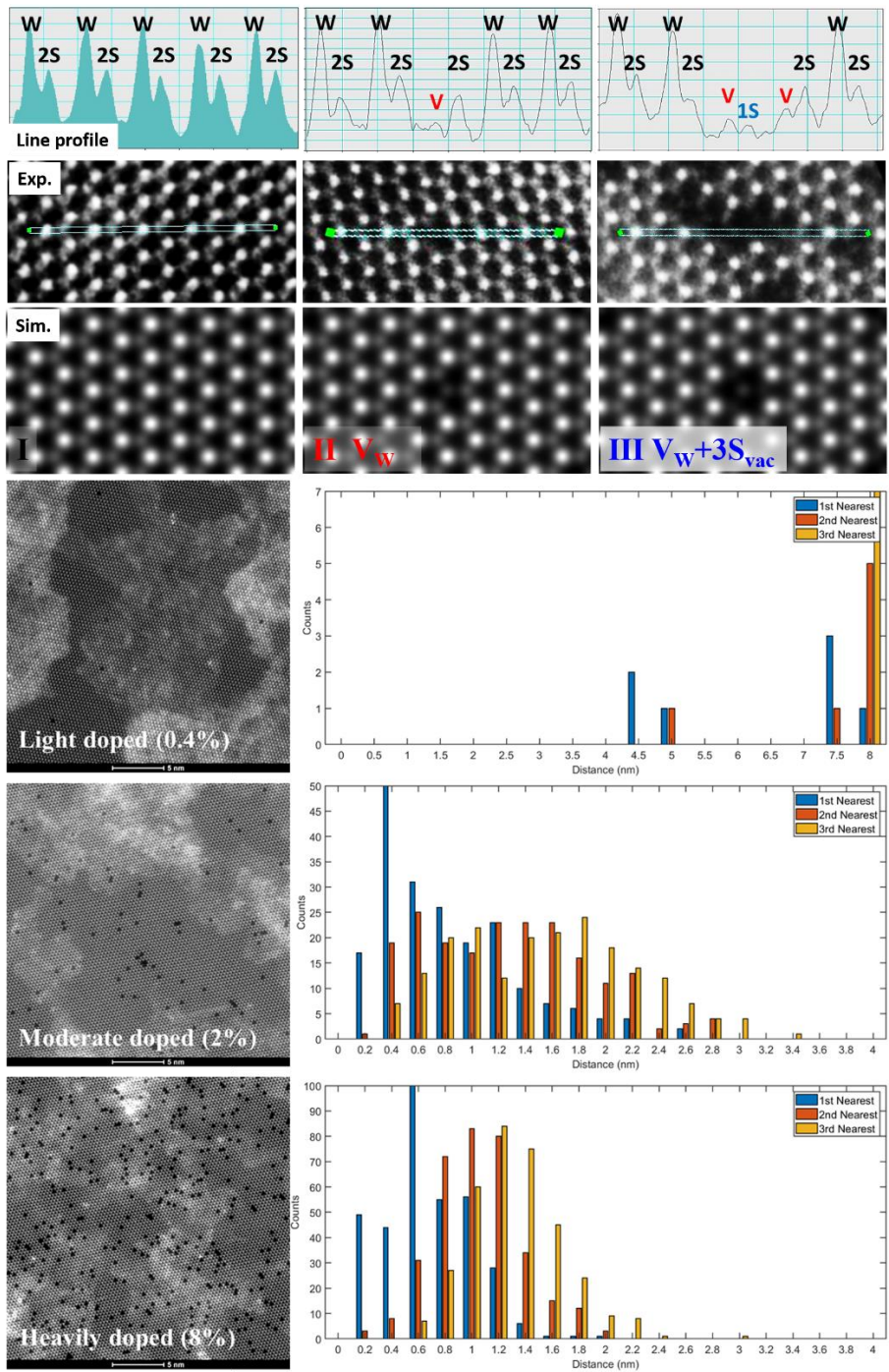

Figure 7 HAADF-STEM images and line profiles of V-doped $WS_2$. Upper: Intensity line profile of the experimental STEM image with pristine $WS_2$ lattice (black curve), $V_W$ and $V_W+S_{vac}$, respectively. Middle: corresponding experimental and simulated STEM images. Bottom: the corresponding histogram of the V-V near-neighbor distance distribution for V-doped $WS_2$ monolayers.

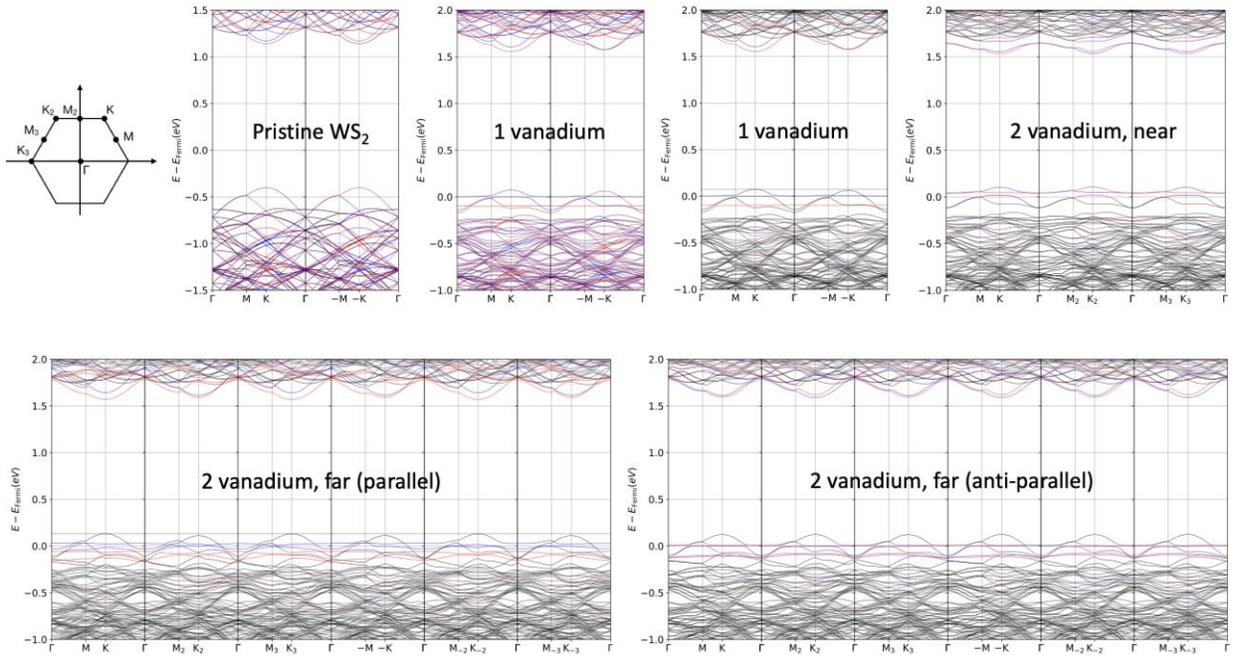

Figure 8 Bandstructures for 7×7 WS$_2$ supercell with different vanadium doping levels and dopant spin textures. The red/blue shows spin-up/down. For bandstructures with only several bands colored, the coloring is the projection onto vanadium $d_{z^2}$ orbital to characterize the defect states. Along M-K-Γ, most bands flip the spin state while the defect states clearly remain the same spin state. The dashed line at the valance band top is to show the band maximum difference between K and –K.

In the bandstructure of pristine WS$_2$, similar to other work,[4] a giant 427 meV spin-orbit splitting at K (–0.401 eV and –0.828 eV) can be seen although it is folded. For the bandstructure of a single vanadium dopant, the spin splitting for the defect state is 0.10 eV, the valence band maximum difference between valleys is 12.6 meV, and the two spin states at the conduction band minimum in the –K valley are almost degenerate. For the bandstructure with 2 vanadium dopants at nearest-neighbor sites, the system converges to no spin polarization state, so both time-reversal and reflection symmetry (the "mirror" is between the two dopants) are preserved. For 2 vanadium dopants sitting farthest from each other with parallel spin directions, the valence band maximum differs by 19.5 meV between valleys and the conduction band minima shift enough to have the same spin state among all valleys.

9. Estimation of the optimal doping level to obtain largest saturation magnetization

Suppose the doping level is $p$, i.e. the likelihood of a given metal site being V is $p$ and $(1-p)$ for W. Starting from a given V dopant site, we assign half of the net magnetic moment given in Table 1 according to nearest dopant neighbor, assuming a random alloy. Specifically, the possibility of a V having its nearest dopant neighbor sitting at the 3$^{rd}$ nearest metal site is $(1-p)^{12}[1-(1-p)^6]$, and we assign 0.46 $\mu_B$ to it. The possibility of its nearest dopant neighbor sitting at larger distance is $(1-p)^{18}$ and we assign a rough averaged moment in this region, 0.64 $\mu_B$ to it. Therefore, the expectation value of the moment contributed by a single V is roughly $0.46(1-p)^{12}[1-(1-p)^6] + 0.64(1-p)^{18}$. The effective moment doping level will then be $p\{0.46(1-p)^{12}[1-(1-p)^6] + 0.64(1-p)^{18}\}$, whose maximum occurs at $p \approx 7\%$ with a corresponding magnetization of $2 \times 10^{-3}$ $\mu_B$/Å$^2$. This simple estimate assumes an ideal random-alloy distribution of vanadium and a magnetic moment contribution based purely on dopant interactions with its nearest neighbor, but it gives a rough upper limit for optimal doping level, since the interaction with more neighbors will likely make the defect states more dispersive and reduce the net moments.

As an illustrative comparison, we present below the magnetic properties calculated without spin-orbit coupling.

Table 1 Energy and net magnetic moment for two vanadium atoms in a supercell (no spin-orbit coupling)

(★) shows the result is not spin-polarized. The antiferromagnetic state has a lower energy when neglecting spin-orbit coupling, and the spin-split dopant levels separate from the valence band sufficiently to attain integral occupancy. The general effect of quenching of the moment at close dopant separations is preserved even in the absence of spin-orbit interaction.

| Dopant pair in lattice coords (Å) | Pair separation (Å) | Energy of the most stable spin texture (meV) | Net magnetic moment ($\mu_B$) | Energy of competing spin texture (meV) | Moment of competing spin texture ($\mu_B$) |
|---|---|---|---|---|---|
| −1 0 | 3.19 | 0 (★) | 0.00 | -- | -- |
| 1 1 | 5.52 | 86.1 (★) | 0.00 | -- | -- |
| 0 2 | 6.38 | 94.1 (★) | 0.00 | -- | -- |
| 1 2 | 8.44 | 123.4 (↑↓) | 0.00 | 144.6 (↑↑) | 2.00 |
| 0 3 | 9.57 | 132.1 (↑↓) | 0.00 | 142.5 (↑↑) | 2.00 |
| 2 2 | 11.02 | 133.1 (↑↓) | 0.00 | 137.4 (↑↑) | 2.00 |
| 1 3 | 11.49 | 132.6 (↑↓) | 0.00 | 137.1 (↑↑) | 2.00 |

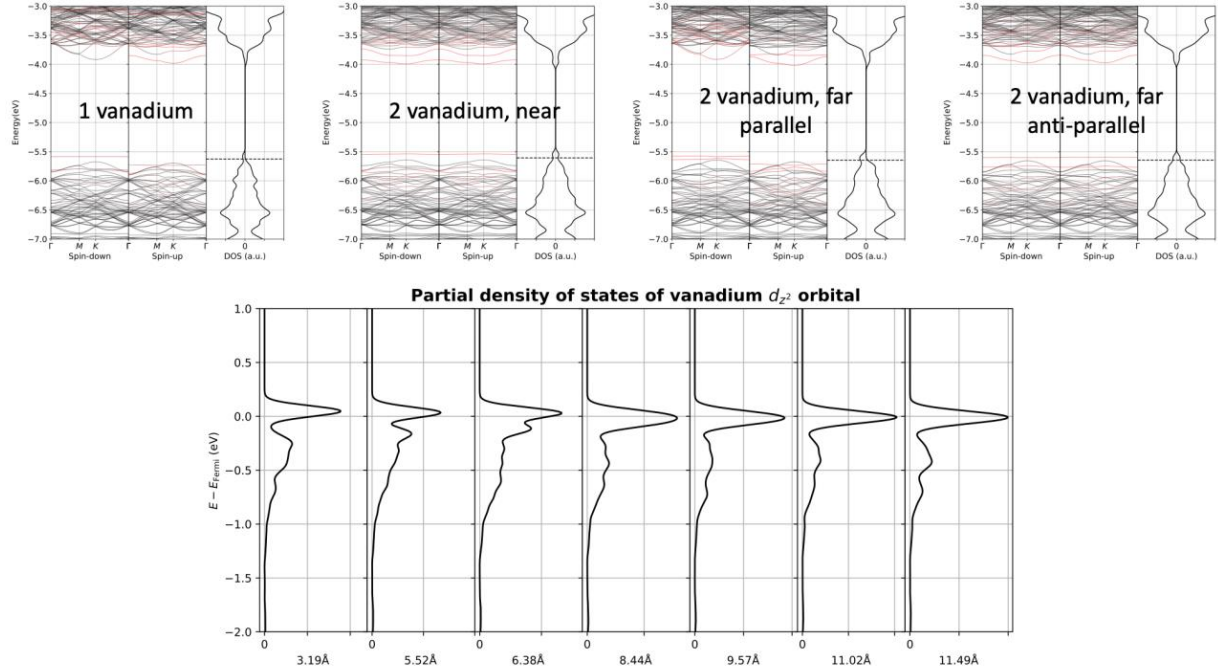

Figure 9 Bandstructures without spin-orbit coupling and the partial density of states without spin polarization. The red color of the bands represents a projection onto vanadium $d_{z^2}$ orbital to characterize the defect states. The vacuum level has been set to 0.

Without spin-orbit coupling, the vanadium is a shallow dopant. From the one vanadium case, the spin splitting (0.138 eV) of the defect state brings one spin state down to the valence band to become fully occupied while the other spin state is fully unoccupied, resulting in an integral Bohr magneton polarization. For the case of two vanadium atoms sitting at nearest-neighbor sites, the system shows no spin polarization. Anti-bonding state is fully unoccupied and the bonding state hybridizes with $WS_2$ valence bands. For 2 vanadium dopants sitting farthest from each other with parallel spin directions, the two spin-down defect states are fully unoccupied, resulting in 2 Bohr magnetons of net magnetization. If they have anti-parallel spins, the two spin branches are nearly degenerate and the up and down spin states are localized at different dopant sites.

The quenching of the magnetic moments in the cases without spin-orbit coupling could be understood by comparing the bonding/anti-bonding splitting and the spin splitting of the defect state. If the bonding/anti-bonding splitting is larger, the moments are quenched. We choose $d_{z^2}$ to track the defect state, since the single vanadium defect state projects only onto the V $d_{z^2}$ orbital (and Mo and S orbitals). The bonding/anti-bonding energy splitting is 0.31, 0.20, and 0.15 eV for the closest three separations, beyond which the splitting is not resolved (<0.1 eV) due to the 0.05 eV Gaussian smearing. For comparison, the spin splitting is 0.138 eV. The pair of bonding/antibonding levels is most apparent at 5.52 and 6.38 Å separations (for 3.19 Å, the lower defect state splits sufficiently that it overlaps with other valence band states).